\documentstyle[11pt,newpasp,twoside,epsf]{article}
\def\edcomment#1{\iffalse\marginpar{\raggedright\sl#1\/}\else\relax\fi}
\reversemarginpar

\begin{document}
\title{VSOP Observations of the EGRET Blazar 1633+382} 
%VSOP Imaging of the $\gamma$-ray Blazar 1633+382}

\author{C.~C. Cheung}
\affil{Department of Physics, Brandeis University, Waltham, MA 02454}
%ccheung@brandeis.edu, MS~057,

\author{J.~S. Ulvestad}
\affil{National Radio Astronomy Observatory, Socorro, NM 87801}
%P.O. Box 0,

\author{W.~T. Vestrand}
\affil{Los Alamos National Laboratory, Los Alamos, NM 87545}
%NIS-2 Division

\author{J.~G. Stacy}
\affil{Louisiana State University, Baton Rouge, LA 70803}

\begin{abstract}
Two high-resolution 5 GHz VSOP+VLBA images of the blazar 1633+382 reveal
clear morphological change in both the bright inner east-west jet, and in
the more diffuse structure further down the jet to the northwest. Here, we
discuss the motion of the inner jet as traced by these VSOP images and
earlier VLBA observations.
\end{abstract}

\section{Background}

The quasar 1633+382 shows a $\gamma$-ray bright blazar nucleus (Mattox et
al. 1993) with a compact mm-VLBI core whose measured brightness
temperature is near the inverse-Compton limit (Krichbaum et al. 2002).
Early VLBI observations showed a superluminal pc-scale jet (Barthel et al.
1995) embedded in an unresolved arcsecond-scale core straddled by two
faint north-south lobes $\sim$14$\arcsec$ in total extent (Murphy, Browne,
\& Perley 1993). Recent VLBA observations (e.g., Fey \& Charlot 1997;
Jorstad et al. 2001) reveal a predominantly E-W core-jet structure, and a
faint extension to the NW which aligns best with the direction of the
emission in which the original superluminal motion was tracked.

\section{VSOP Imaging of Superluminal Motion in the Jet}

Our 5 GHz images of 1633+382 clearly separate the VLBI radio source into
two almost equally bright components -- the core to the east and the jet
extending to the west (Fig.~1). Much more substructure is apparent in the
second epoch VSOP observation where much greater maximum projected
baselines were reached with the spacecraft, matching the resolution of
ground only observations at higher frequencies (e.g. at 22 GHz, Jorstad et
al. 2001). This second epoch data (which also used the full VLA) was
shared with Lister et al. (2001)  as part of their imaging of the
Pearson-Readhead sample with VSOP, and our full resolution image is very
similar to their published image so is not shown here.

Our proper motion analysis is limited by the resolution achieved in the
first epoch VSOP observation.  We modelfitted our data in the u-v plane
with DIFMAP, and in the image plane with AIPS assuming a simple double
model to fit the core and inner jet component. The best fit positions of
the secondary relative to the core are 1.37 mas (July 1997) and 1.56 mas
(August 1998) at a position angle (PA) of $-85\deg$. At higher resolution
(Jorstad et al. 2001; Lister et al. 2001), these two features are resolved
further making the distance between the most prominent peak in the jet and
the core in their maps somewhat greater.

As part of a separate program (Ulvestad et al.), we obtained an earlier
(1995) VLBA 8.4 GHz image of 1633+382 (Fig.~1) and can identify the same
two inner components as in our VSOP 5 GHz maps, and measured their
separation to be 1.01 mas at a PA=$-85\deg$, consistent with results found
by Fey \& Charlot (1997) at a nearby epoch. From a 1996 VLBA 5 GHz
observation, Fomalont et al. (2000) found a separation of 1.2 mas at the
same PA.

Taking these low frequency data together, we can infer a fairly constant
motion of about 0.17 mas/yr along a PA=$-85\deg$ over the period between
1995 and 1999. At z=1.814, this corresponds to $v_{\rm app}\simeq13$c
(H$_{0}=70~$km~s$^{-1}$~Mpc$^{-1}$, $\Omega_{\rm M}=0.3$ and
$\Omega_{\Lambda}=0.7$). This implies that the jet is aligned at
$\leq9\deg$ to our line of sight and the Lorentz factor corresponding to
the pattern velocity is $\Gamma=13$ or greater. The epoch of zero
separation is thus about 1989.2.  This is the same zero epoch of
separation inferred from the higher resolution VLBA monitoring of the
compact edge of the same dominant jet feature at 22 GHz by Jorstad et al.  
(2001) where they measured a motion of 0.2 mas/yr also along the E-W
direction.  For comparison, the early ejection found by Barthel et al.
moved 0.16 mas/yr at a more northerly (15$\deg$ to 25$\deg$) projected PA
in the sky. Future observations could determine if this current strong
long lasting feature will continue in its current E-W trajectory or follow
the path of the older emission.

\acknowledgements
The NRAO is operated by Associated Universities, Inc., under cooperative
agreement with the NSF. Space VLBI at NRAO is funded by NASA. The VSOP
project is led by the Japanese ISAS in cooperation with many organizations
and radio telescopes around the world. Radio astronomy at Brandeis is
supported by the NSF.

\begin{figure}
%\epsscale{1.1}
\vspace{-0.65in}
\plotone{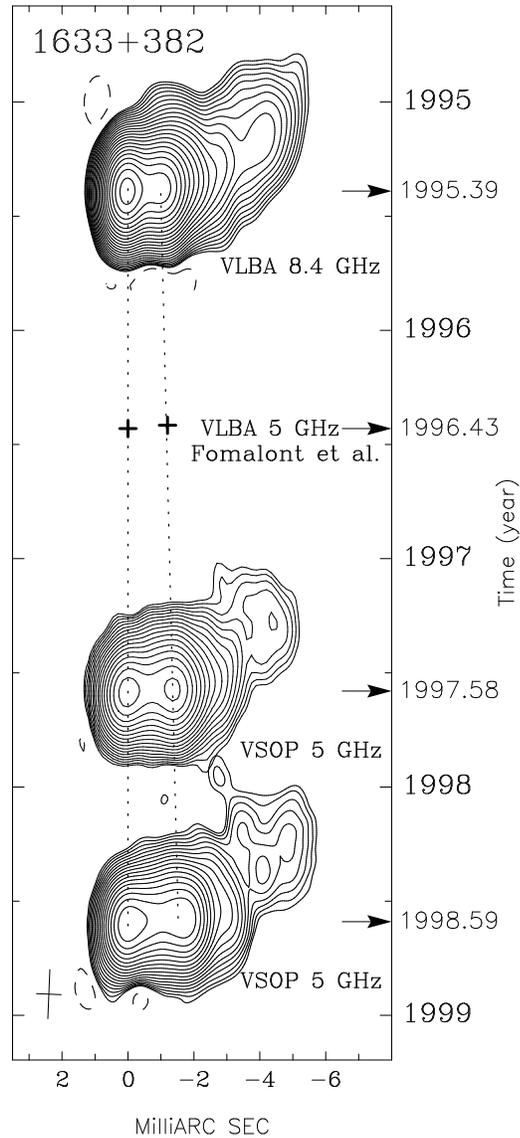}
\vspace{-0.4in}
\caption{
Sequence of VLBI images of 1633+382 convolved with a common beam (bottom
left corner) of $1.49\times 0.75$~mas at PA$=-2.42\deg$ (achieved by the
1997 VSOP data).  Contours are spaced by factors of $\sqrt{2}$ with
[min, max] levels of [0.4, 819.2] mJy/bm (8.4 GHz image), and [2, 512] mJy/bm (VSOP images).  
Map peaks are (top to bottom)
0.995, 0.647, and 0.718~Jy/bm. Arrows indicate the observation epochs.
Positions of the core and jet
component from Fomalont et al. (2000) are indicated with plus signs. The dotted lines trace the 0.17
mas/yr ($\sim$13c) motion of the western jet feature relative to the
presumed stationary core.
}
%\plotfiddle{q1633_3panel_Plus.eps\hsf{50}}   
\end{figure}

\end{document}